\documentclass{article}
\usepackage{frascatiphys_C}
\usepackage{graphicx}

\begin{document}
\title{ PRECISION SPECTROSCOPY OF PIONIC ATOMS: 
FROM PION MASS EVALUATION 
TO TESTS OF CHIRAL PERTURBATION THEORY}
\author{Martino Trassinelli \thanks{\hspace{0.2cm} On behalf of the PIONIC HYDROGEN and PION MASS collaboration\cite{proposals}}\\
\em   Laboratoire Kastler Brossel,Universit\'e P. et M. Curie, F-75252 Paris, France
}
\maketitle
\baselineskip=11.6pt
\begin{abstract}

Preliminary results of the strong interaction shift and width 
 in pionic hydrogen ($\pi H$) using an X-ray spectrometer with 
spherically bent crystals and CCDs as X-ray detector are presented. In the experiment at the Paul Scherrer Institute 
three different $(np\to1s)$ transitions in $\pi H$ were measured.
Moreover the pion mass measurement using the $(5 \to 4)$ transitions in
pionic nitrogen and muonic oxygen is presented
\end{abstract}

\section{Introduction}
Pionic hydrogen atoms are unique systems to study the strong interaction at low energies\cite{gotta2004}.  
The influence of the strong interaction in pionic hydrogen can be extracted from the $(np\to1s)$ transitions. 
Compared to pure electromagnetic interaction, the 1s level is affected 
by an energy shift $\epsilon_{1s}$ and a line broadening $\Gamma_{1s}$. The shift and the broadening are related to the hadronic scattering lengths $a^h_{\pi^-\ p  \to \pi^-\ p }$ and $a^h_{\pi^-\ p  \to \pi^0\ n }$, by the Deser-type formulae \cite{deser}:

\begin{equation}
\frac{\epsilon_{1s}}{E_{1s}}=-4\frac{1}{r_B}a^h_{(\pi^- \ p \to \pi^- \ p)}(1+\delta_\epsilon) \label{eq:deser1}
\end{equation}

\begin{equation}
\frac{\Gamma_{1s}}{E_{1s}}=8 \frac{Q_0}{r_B}(1+\frac{1}{P})(a^h_{(\pi^- \ p \to \pi^0 \ n)}(1+\delta_\Gamma))^2 \label{eq:deser2}
\end{equation}
where $\epsilon_{1s}$ is the strong interaction shift of the 1s level reflecting the $\pi\,p$ scattering process.
$\Gamma_{1s}$ is the width of the ground state caused by the 
reactions $\pi^- + p \to \pi^0  + n$ and $\pi^- + p \to \pi^0  + \gamma $.
$Q_0=0.1421~fm^{-1}$ is the kinetic center of mass momentum of the $\pi^0$ in $\pi^- + p \to \pi^0  + n$ reaction, 
and $P=1.546 \pm 0.009$\cite{spuller1977}  is the branching ratio of the charge exchange and radiative capture (Panofsky ratio).
$\delta_{\epsilon,\Gamma}$ are corrections that permit to connect the pure
hadronic scattering lengths to the measurable shift and width\cite{gasser2002,sigg1996th,ericson2003}.
The hadronic scattering lengths can be related to the isospin-even and isospin-odd scattering length,
$a^+$ and $a^-$:

\begin{equation}
a^h_{(\pi^- \ p \to \pi^- \ p)}= a^+ + a^-\qquad a^h_{(\pi^- \ p \to \pi^0 \ n)}=-\sqrt{2}\ a^-
\end{equation}

The isospin scattering lengths can ben related to $\epsilon_{1s}$ and $\Gamma_{1s}$ in the framework of the 
 Heavy Barion Chiral Perturbation Theory ($\chi$PT)\cite{lyubovitskij2000}.
Scattering experiments are restricted to energies above 10~MeV
and have to rely on an extrapolation to zero energy to extract the scattering lengths. 
Pionic hydrogen spectroscopy permits to measure this scattering length at almost zero energy 
(in the same order as the binding energies, i.e., some keV) and verify with high accuracy the $\chi$PT calculations. 
Moreover, the measurement of $\Gamma_{1s}$ allows an evaluation of the pion-nucleon coupling constant $f_{\pi N}$, which is related to $a^-$ by the Goldberger-Miyazawa-Oehme sum rule (GMO)\cite{GMO}.

Pionic atom spectroscopy permits to measure another important quantity: the charged pion mass.
Orbital energies of pionic atoms depend on the reduced mass of the system.
These energies can be calculated with high accuracy using Quantum Electrodynamics.
Measuring transition energies, not disturbed by strong interaction, allows to determine the reduced mass of the system and hence the mass of the pion.
The accurate value of the pion mass is crucial to evaluate the upper bound of the mass of the muonic neutrino from a measurement of the pion decay\cite{nelms2002}.

\section{Description of the setup}
The pionic atoms are produced using the pion beam provided by the Paul Scherrer Institut\cite{proposals}. The beam momentum is 110~MeV/c with an intensity of $10^{8}~{\rm s}^{-1}$.
The pions are captured and slowed down using a cyclotron trap\cite{simons1993}.
The target is made of a cylindrical cell with Kapton walls, positioned in the center of the trap.
In the target cell the decelerated pions are captured in bound atomic states. During the de-excitation X-rays are emitted. 
As the muons from the pion decay in the beam are present as well, it is possible to produce muonic atoms and pionic atoms at the same time.
The X-ray transition energies are measured using a bent crystal spectrometer and a position sensitive detector.
The reflection angle $\Theta_B$ between the crystal planes and the X-rays is related to the photon wavelength $\lambda = h c / E$ 
by the Bragg formula:
\begin{equation}
n\ \lambda = 2\ d \sin \Theta_B
\end{equation}
where {\it n} is the order of the reflection and $d$ is the spacing of the crystal planes.\\
The detector is formed by an array of 6 CCDs composed each by $600 \times 600$ pixels\cite{nelms2002}, the pixel size is $40~\mu m$. 
The 3-4~keV X-rays excite mostly one 
or two pixels. Larger clusters are due to charged particle or high-energy gamma radiation and can be eliminated by cluster analysis. Transitions of different energies result in different reflection lines on the detector. By measuring the distance between these lines it is possible to determine the energy difference.
The resolution of the spectrometer is of the order of $0.4~eV$ at 3~keV.

\section{Extraction of the hadronic shift and width}
\begin{figure}[t]
\begin{center}
\includegraphics[width=0.47\textwidth]{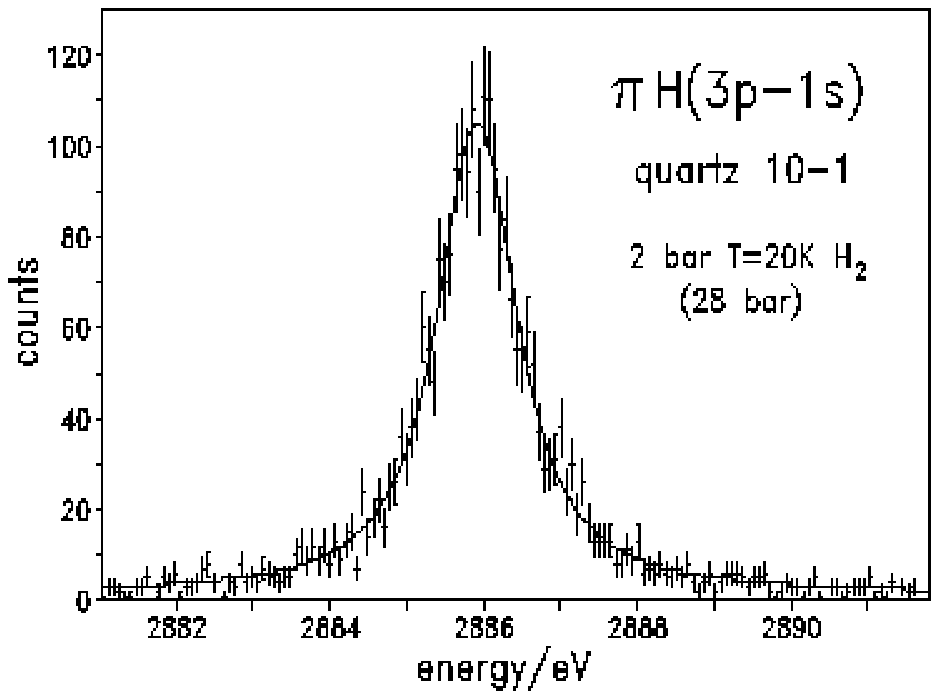} \hspace{0.2cm}
\includegraphics[width=0.47\textwidth]{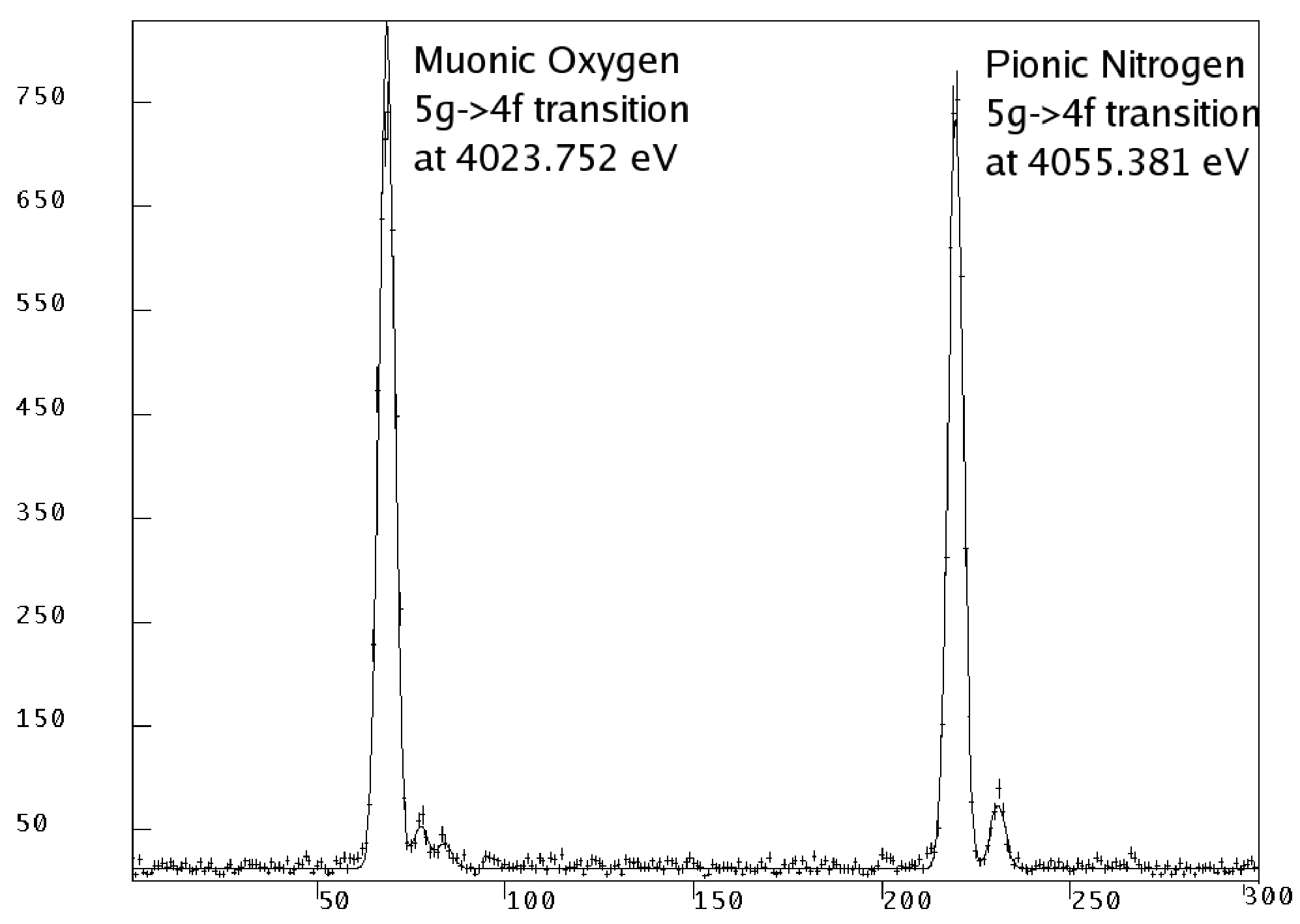}
\end{center}
\caption{\it Left: $3p \to 1s$ transition measurement of pionic hydrogen. Right: $5g \to 4f$ transition in pionic oxygen and muonic hydrogen.} \label{spectra} 
\end{figure}

The characteristics of the ground state of pionic hydrogen are evaluated measuring the X-ray transitions $np \to 1s$ (see fig.\ref{spectra}).
The line width is the result of the convolution of: the spectrometer resolution, the Doppler broadening effect from the
non-zero atom velocity, the natural width of the ground state, and, of course, the hadronic broadening.
A very accurate measurement of the response function of the crystal was performed using the $1s2s\,^{3}S_{1}\to1s^{2}\,^{1}S_{0}$ M1 
 transitions in He-like argon (with a natural line width less than 1 meV, Doppler broadening about 40 meV). 
For this measurement the cyclotron trap was converted into an Electron-Resonance Ion Trap 
(ECRIT)\cite{anagnostopoulos2003}, with the crucial point that the geometry of the setup was preserved.\\
The Doppler broadening effect in the pionic transitions can be studied by working at different pressures and with different transitions. With the help of a cascade model we can predict
the kinetic energy distribution of the atoms and the corresponding Doppler 
broadening\cite{jensen2002a}.\\ 
A first series of measurements were completed in 2002. 
The hadronic broadening $\Gamma_{1s}$ extracted from the experimental line width is:

\begin{equation}
\Gamma_{1s}= 0.80 \pm 0.03~eV 
\end{equation}

By varying the target density, we were able to prove that the formation of complex systems $\pi p + H_2 \to [(\pi p p ) p ] e e$\cite{jonsell1999}, which can add an additional shift to the ground state, is negligible.
Energy calibration for the $\pi H(3p \to 1s)$ transition is performed using the
$6h \to 5g$ transition in pionic oxygen. Strong interaction
and finite nucleus size effect are negligible for this transition.
Orbital energies can be calculated with an accuracy of a few meV\cite{paul}.
The result for the shift is:

\begin{equation}
\epsilon_{1s}=7.120 \pm 0.017~eV
\end{equation}

For the calculation of the shift a pure QED value of $E^{QED}_{3p-1s}=2878.809~eV$ was used. The above given errors include statistical accuracy and 
systematic effects \cite{maik2003}. The value of  $\epsilon_{1s}$ is in agreement with the result
of a previous experiment, where the energy calibration was performed with $K \alpha$ fluorescence X-rays\cite{schroder2001},
but more precise by a factor of 3.
\section{Pion mass measurement}
\begin{figure}[t]
\begin{center}
\includegraphics[height=0.5\textwidth]{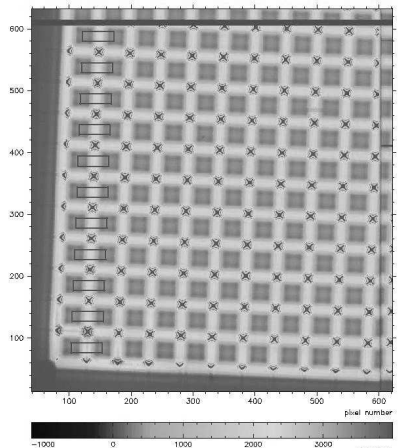}
\end{center}
\caption{\it Detail of the grid image on the CCD detector with the selected zones for the linear fit.
} \label{mask} 
\end{figure}

The evaluation of the pion mass is obtained by the measurement of the transition energy 
 of the $5g \to 4f$ transition in pionic nitrogen in 2000 (see fig.\ref{spectra}). 
We used the analog transition in 
muonic oxygen as a reference line. The energy difference between the two lines depends on the ratio between 
the pion mass and the muon mass, which is known with 0.05~ppm accuracy. The expected accuracy
for the pion mass is less than 2~ppm, to be compared with the actual value, 
which has an accuracy of 2.5~ppm. This value is the average of two measurements, obtained using two different techniques and which differ by 5.4~ppm\cite{PDG2004}.
To reach this precision, we need a perfect understanding of the crystal spectrometer aberrations
and the exact distance between pixels in the detector.
For the second task an experiment was set up in September 2003 to measure the 
pixel distance at the working temperature of $-100^\circ$C. 
We used a nanometric grid composed by $21 \times 14$ lines, $20~\mu m$ thick, spaced by 2~mm with an accuracy of about $0.05~\mu m $.
The mask, at room temperature, was illuminated by a point-like source at a distance of 6426.7~mm, and positioned at
37~mm from the CCD detector (see fig.\ref{mask}).
Applying linear fits to the lines of the grid  in the CCD image it was possible to
provide an accurate measurement of the average pixel distance:

\begin{equation}
pixel\ distance = 39.9943 \pm 0.0035~\mu m
\end{equation}

\section{Conclusions and outlook}
The strong interaction shift in pionic hydrogen has been determined with an accuracy of 0.2\%.
During spring-summer 2004 the crystals have been characterized with X-rays from the ECRIT.
The measurement of the broadening in muonic hydrogen in November-December 2004, together with the cascade model, 
will allow us to reach an accuracy of 1\% for $\Gamma_{1s}$.

\section{Acknowledgments}
We thank the PSI staff, in particular the nanotechnology group, which provided
the nanometric mask for the pixel measurement.

\end{document}